# Mie Scattering of Phonons by Point Defects in IV-VI Semiconductors PbTe and GeTe

Ruiqiang Guo[1] and Sangyeop Lee[1,2,*]

[1]*Department of Mechanical Engineering and Materials Science, University of Pittsburgh, Pittsburgh, Pennsylvania 15261, USA*

[2]*Department of Physics and Astronomy, University of Pittsburgh, Pittsburgh, Pennsylvania 15261, USA*

\* sylee@pitt.edu

**Abstract**

Point defects in solids such as vacancy and dopants often cause large thermal resistance. Because the lattice site occupied by a point defect has a much smaller size than phonon wavelengths, the scattering of thermal acoustic phonons by point defects in solids has been widely assumed to be the Rayleigh scattering type. In contrast to this conventional perception, using an *ab initio* Green's function approach, we show that the scattering by point defects in PbTe and GeTe exhibits Mie scattering characterized by a weaker frequency dependence of the scattering rates and highly asymmetric scattering phase functions. These unusual behaviors occur because the strain field induced by a point defect can extend for a long distance much larger than the lattice spacing. Because of the asymmetric scattering phase functions, the widely used relaxation time approximation fails with an error of ~20% at 300K in predicting lattice thermal conductivity when the vacancy fraction is 1%. Our results show that the phonon scattering by point defects in IV-VI semiconductors cannot be described by the simple kinetic theory combined with Rayleigh scattering.

## 1. Introduction

Scattering of waves by the inhomogeneity of the medium is a fundamental process underpinning diverse applications including electromagnetics, optics, and acoustics [1-4]. This is of equal importance for thermal phonons [5-7]. Randomly distributed particles can result in phonon localization through multiple scattering and interference of phonon waves [8,9]. Periodic inclusions in phononic crystals give rise to new vibrational band structures due to Bragg and Mie scatterings [10,11]. The wave-defect interaction is usually assumed elastic and its characteristics dramatically depend on the size of the scattering centers. The elastic scattering of phonons, analogous to that of electromagnetic (EM) waves, falls into three regimes: (i) Rayleigh scattering for the scattering centers much smaller than the phonon wavelength ($d \ll \lambda$ where $d$ and $\lambda$ are the diameter of the scattering center and the wavelength, respectively), (ii) Mie scattering for those with comparable size to the wavelength ($d \sim \lambda$), and (iii) Geometric scattering for those much larger than the wavelength ($d \gg \lambda$) [12,13].

A point defect at a single lattice site possesses the minimum size of a scattering center in solids. Because the wavelengths of thermal phonons are usually many times of the lattice spacing, the scattering of thermal phonons by a point defect is often considered Rayleigh scattering and exhibits very weak scattering strength for low frequency phonons. The Rayleigh scattering is well captured by the models suggested by Klemens [14] and Tamura [15], which assume that the change in mass and interatomic force constants (IFCs) by a point defect is limited to a single lattice site. The recently developed *ab initio* Green's function approach [16,17] also shows the Rayleigh scattering behavior for strongly bonded materials, such as diamond and boron arsenide (BAs) [18-24].

In this report, we show that the point defects in IV-VI semiconductors (PbTe and GeTe), unlike the previously studied materials, cause Mie scattering of thermal phonons. The IV-VI semiconductors are widely used for the applications of thermoelectrics and phase change materials where the thermal conductivity is a key parameter for the device figure-of-merit. Particularly, large amounts of vacancies (1~3%) have been often reported even in single crystals of these materials, which can reduce the lattice thermal conductivity by half or more [25-36]. Also, it is common that thermoelectric materials contain large concentrations of substitutional defects (1 to 2%) to achieve an optimal doping level for a large thermoelectric power factor [26-31].

## 2. Computational method

*2.1 Phonon scattering rates by point defects*

*2.1.1 Green's function approach*

When a phonon mode $\mathbf{q}s$ ($\mathbf{q}$ and s represent the wavevector and polarization, respectively) is scattered into another phonon mode $\mathbf{q}'s'$ elastically by a point defect, the transition rate can be calculated using the Green's function approach [16-24]:

$$\Gamma^{d}_{\mathbf{q}s,\mathbf{q}'s'} = \frac{\pi\Omega}{\omega V_d}\left|\langle\mathbf{q}'s'|\mathbf{T}^{+}(\omega^2)|\mathbf{q}s\rangle\right|^2 \delta(\omega^2_{\mathbf{q}'s'} - \omega^2_{\mathbf{q}s}), \tag{1}$$

where $\omega$ is the angular frequency of phonons, $\Omega$ is the unit cell volume, $|\mathbf{q}s\rangle$ is the phonon eigenstate, and $V_d$ is the defect volume. The $\mathbf{T}$ matrix is defined as

$$\mathbf{T}^{+} = (\mathbf{I} - \mathbf{V}\mathbf{G}_0^{+})^{-1}\mathbf{V}, \tag{2}$$

where $\mathbf{I}$ is the identity matrix, $\mathbf{V}$ is the perturbation matrix representing the differences of atomic mass and IFCs induced by the point defect. The $\mathbf{G}_0^{+}$ is the retarded Green's function of the unperturbed crystal, which is calculated numerically using the tetrahedron approach [37]. In the general case, the perturbation matrix $\mathbf{V}$ can be decomposed into $\mathbf{V}^M$ and $\mathbf{V}^K$, which represent the changes of atomic mass and IFCs, respectively. The $\mathbf{V}^M$ is expressed as

$$\mathbf{V}^M = -\frac{M'_i - M_i}{M_i}\omega^2, \tag{3}$$

where $M_i$ and $M'_i$ are the atomic masses at the lattice site $i$ in the perfect and defected crystals, respectively. The $\mathbf{V}^K$ is constructed based on the IFCs of the perfect ($K$) and defected ($K'$) structures, with each matrix element calculated by

$$\mathbf{V}^K_{i\alpha,j\beta} = \frac{K'_{i\alpha,j\beta} - K_{i\alpha,j\beta}}{(M_i M_j)^{1/2}}, \tag{4}$$

where $\alpha$ and $\beta$ are Cartesian axes, $M_i$ and $M_j$ represent the pristine atomic masses at lattice sites $i$ and $j$, respectively. The perturbation matrix consists of only $\mathbf{V}^K$ for a vacancy but both $\mathbf{V}^K$ and $\mathbf{V}^M$ for substitutional defects.

Assuming the scattering of phonons by a point defect is independent of other point defects, the phonon-defect scattering rate is

$$\frac{1}{\tau_{\mathbf{q}s}^d} = f \sum_{\mathbf{q}'s'} \Gamma_{\mathbf{q}s,\mathbf{q}'s'}^d , \tag{5}$$

where $f$ is the volumetric fraction of point defects.

*2.1.2 Born approximation and the Klemens' model using realistic phonon dispersion*

For weak perturbations, the aforementioned calculation can be simplified using the Born approximation, i.e., substituting the **T** matrix in Eq. (1) with the perturbation matrix **V**, which is the first-order approximation of the Born series:

$$\mathbf{T}^+ = \mathbf{V} + \mathbf{V}\mathbf{G}_0^+\mathbf{V} + \mathbf{V}\mathbf{G}_0^+\mathbf{V}\mathbf{G}_0^+\mathbf{V} + ... \approx \mathbf{V} . \tag{6}$$

Klemens suggested a simple model for the rate of phonon-vacancy scattering based on the Debye phonon dispersion and an effective mass variance $(\Delta M / M)^2 = 9$ [38]. To use the realistic phonon dispersion from *ab initio* calculation, we combine the Klemens' effective mass variance with the Tamura's model for phonon-isotope scattering:

$$\frac{1}{\tau_{\mathbf{q}s}^{iso}} = \frac{\pi \omega_{\mathbf{q}s}^2}{2N} \sum_{\mathbf{q}'s'} \sum_{i} g(i) \left| \mathbf{e}_{\mathbf{q}s}^*(i) \cdot \mathbf{e}_{\mathbf{q}'s'}(i) \right|^2 \delta(\omega_{\mathbf{q}s} - \omega_{\mathbf{q}'s'}) , \tag{7}$$

with the **q**-point mesh size $N$. The mass variance $g(i)$ is defined as $\sum_b f_b(i)[1 - M_b(i)/\bar{M}(i)]^2$ where $f_b(i)$ and $M_b(i)$ are the fraction and mass of the $b$th isotope of the atom $i$, and $\bar{M}(i)$ is the average mass of the $i$th atom in the unit cell. In this work, the Klemens' model for vacancy cases refers to Eq. (7) with the Klemens' effective mass variance for $g(i)$.

*2.2 Density functional theory calculation*

We determined the harmonic and anharmonic IFCs from density functional theory (DFT) calculations using the projector augmented wave method [39], as implemented in the Vienna *ab initio* simulation package (VASP) [40]. The Perdew-Burke-Ernzerhof (PBE) generalized gradient approximation was used for the exchange-correlation functional for all materials [41]. In Table S1, we list the cutoff energy and **k**-point sampling grid for electronic states used in the relaxation of the primitive cell, and the relaxed lattice constants. Also, the spin-orbit interaction was included for the heavy elements Pb and Te. The self-consistent calculation was performed with the force convergence criteria of $10^{-6}$ eV/ Å. The harmonic and anharmonic IFCs were calculated using a supercell of 4×4×4 for Si and a relatively larger one (5×5×5) for PbTe and GeTe in which the long-range interatomic interactions are expected. We test the convergence of IFCs with respect to supercell size, as shown in Fig. S1 in the Supplementary Material.

*2.3 Lattice thermal conductivity*

We calculated the lattice thermal conductivity using two different methods: full matrix and relaxation time approximation (RTA). For the full matrix case, we obtain the total scattering matrix by adding phonon-phonon, phonon-isotope, and phonon-vacancy scattering matrices. For the RTA case, only diagonal terms from the phonon-vacancy scattering matrix are included while full matrices of phonon-phonon and phonon-isotope scattering are used. Then, the linearized Peierls-Boltzmann transport equation was solved with an iterative manner as implemented in ShengBTE [42]. More technical details can be found in previous publications [42-46]. The sampling grids for phonon states in the first Brillouine zone are listed in Table S1. The isotopes of each element are assumed naturally occurring cases. Recent studies show that the phonon renormalization and four-phonon scattering are important for predicting thermal properties of materials such as PbTe, particularly at high temperatures [47-51]. As our study focuses on the Mie scattering of phonons by defects, we ignore the phonon renormalization effect and four-phonon scattering for the sake of simplicity.

**3. Results and discussion**

We use an *ab initio* Green's function approach to calculate the scattering matrix of phonons by point defects in PbTe (space group *Fm3m*) and GeTe (*R3m*) crystals. For comparison, we also present results for Si. The scattering rates were calculated using three different approaches: (i) exact Green's function calculation based on the **T**-matrix formalism, (ii) Green's function calculation to the first order (Born approximation), and (iii) the Klemens' model with the effective mass variance $(\Delta M / M)^2 = 9$ (the 3*M* mass difference is derived by considering the removed kinetic and potential energies due to the removal of an atom from the vacancy site) for single vacancy case [38] combined with Tamura's formula to consider full phonon dispersion [15]. We call the last approach the Klemens' model in this article. The **T**-matrix method and the Born approximation consider mass variance and finite size effects of scattering center by including the changes of IFCs in all nearby atoms, while the Klemens' model assumes an infinitesimally small point defect. Both the Born approximation and the Klemens' model are based on the lowest-order perturbations while the **T**-matrix method includes all high-order terms. Recent *ab initio* Green's function calculations show the failure of the lowest-order perturbative approaches for some strongly bonded materials [18-24].

The calculated phonon-vacancy scattering rates for PbTe, GeTe, and Si are shown in Fig. 1. The Born approximation gives similar results to the **T**-matrix for PbTe and GeTe, while it severely underestimates the scattering rate for Si. This is due to the relatively small perturbations resulting from the weaker bonding in PbTe and GeTe, making the high-order terms in Eq. (6) become less important compared with the case of Si. Illustrating this, the Frobenius norm of self-interaction IFCs for PbTe (Pb site, 3.76 eV/Å$^2$) and GeTe (Ge site, 5.96 eV/Å$^2$) are much smaller than that for Si (22.23 eV/Å$^2$). It is more noteworthy that the frequency dependence of scattering rates of PbTe and GeTe is much different from that of Si. In general, the phonon scattering by defects follows the Rayleigh scattering in the long wavelength limit and the Mie scattering in the short wavelength limit. The Rayleigh scattering model, combined with the density-of-states varying with $\omega^2$ in the long wavelength where $\omega$ is a phonon frequency, results in the typical $\omega^4$ dependence of phonon-vacancy scattering rates. For Si in Fig. 1, this $\omega^4$ dependence at the low frequency range is well captured by all the three approaches. However, for PbTe and GeTe in the same figure, the scattering rates from the **T**-matrix and Born approximation show substantially weaker frequency dependence ($\omega^{2.5}$ and $\omega^3$, respectively) down to 0.1 THz. These results show that the phonon-vacancy scattering in PbTe and GeTe is better described by the Mie scattering rather than the Rayleigh scattering in the full phonon spectrum above at least 0.1 THz.

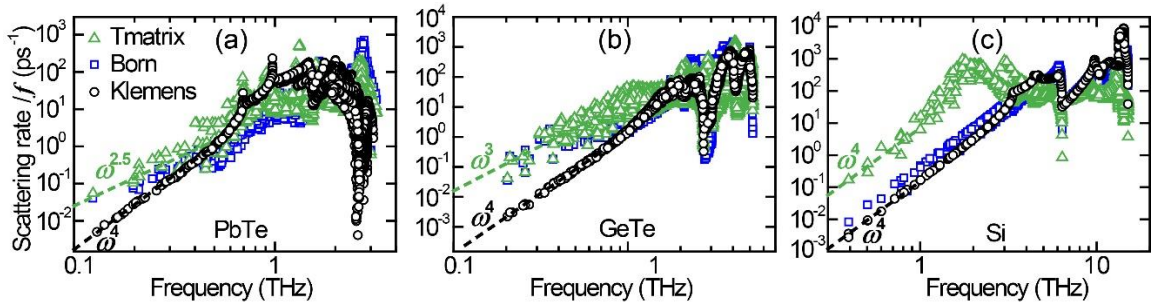

Fig. 1. Phonon-vacancy scattering rates normalized by the atomic fraction of vacancy for (a) PbTe (Pb vacancy), (b) GeTe (Ge vacancy) and (c) Si using three different approaches: **T**-matrix, Born approximation, and the Klemens' model.

Considering that the phonon wavelength corresponding to 0.1 THz (78.8 Å for the transverse acoustic phonon in PbTe) is much larger than the lattice spacing (3.29 Å for PbTe), our observation of the Mie scattering implies that long-range strain field near the vacancy site is more important for phonon scattering in PbTe and GeTe than the vacancy site itself. We evaluate how far the strain field induced by a single vacancy can persist in these materials. We calculated the Frobenius norm of changes in the self-interaction IFCs for each atom $i$, $\|\mathbf{V}_{ii}\|$ where $\mathbf{V}_{ii}$ is the change of the self-interaction IFC tensor, which reflects the overall change of the IFCs and strain field on atom $i$ [22]. Fig. 2(a) shows the $\|\mathbf{V}_{ii}\|$ normalized by $\|\mathbf{V}_{00}\|$ (where 0 denotes the vacancy site) with respect to the distance between the atom $i$ and the vacancy site. All three materials exhibit $\|\mathbf{V}_{ii}\|$ rapidly decreasing for the first two nearest neighbors. Beyond the second nearest neighbors, PbTe and GeTe still have a significantly large $\|\mathbf{V}_{ii}\|$ as marked by red circles while $\|\mathbf{V}_{ii}\|$ of Si is close to zero. Similarly, the calculated displacement of equilibrium position of atoms also indicates long-range strain field in PbTe and GeTe, as shown in Fig. S2 in the Supplementary Material. The long-range strain field induced by a single vacancy in both materials is consistent with the long-range interatomic interactions shown in Fig. 2(b) that were previously reported for perfect crystalline phase [52]. This long-range interaction is due to resonance or hybridization between different electronic configurations, i.e., the unsaturated covalent bonding by $p$-electrons for rocksalt-like crystals [52]. From Fig. 2(a), the changes of IFCs from the vacancy site still exist in the distance of ~ 10 Å, resulting in the strain field with a diameter of at least 20 Å. Thus, thermal phonons with comparable wavelengths experience Mie scattering rather than Rayleigh scattering. This can explain the weaker frequency dependence of phonon-vacancy scattering for those materials shown in Figs. 1(a) and 1(b). In contrast, the $\|\mathbf{V}_{ii}\|$ in Si is short-range, resulting in the $\omega^4$ dependence of the Rayleigh scattering in Fig. 1(c).

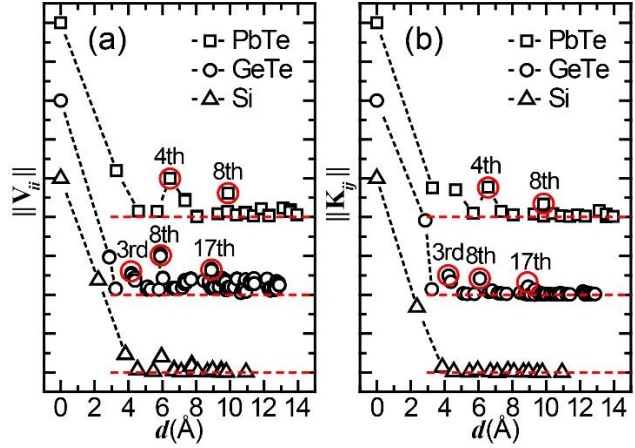

Fig. 2. Interatomic force constants showing the long-range interaction in PbTe and GeTe: (a) Frobenius norm of the change in self-interaction IFCs as a function of distance from the vacancy for PbTe (Pb vacancy), GeTe (Ge vacancy), and Si. (b) Frobenius norm of IFCs in perfect crystals as a function of the interatomic distance measured from Pb site for PbTe and Ge site for GeTe. Note that all the values of $\|\mathbf{V}_{ii}\|$ and $\|\mathbf{K}_{ij}\|$ are normalized by its maximum at $d=0$. The plots are shifted along $y$-direction and the red dashed lines represent $y=0$. The $y$-axes are in linear scale.

A notable feature of Mie scattering of EM waves by a particle is the Mie resonance when the diameter of the particle is an integral multiple of the incident EM wavelength. For our case of phonon scattering in PbTe and GeTe, however, we observe no periodic peaks in the scattering rate versus the wavelength of incident phonons for all three acoustic branches as shown in Fig. 3. The Mie resonant scattering does not exist in our case because the strain field smears out in space and its spatial size cannot be explicitly defined.

Another characteristic of Mie scattering that further distinguishes it from Rayleigh scattering can be found in the scattering phase function, which describes the angular probability distribution of phonon states after scattering from a given incident phonon [2,53,54]. The scattering phase function is typically plotted in a closed shape, with the distance of each point from the center representing the relative likelihood of scattering into that particular direction. Mie scattering usually exhibits a highly asymmetric phase function with forward scattering being more significant than backward scattering, while the phase function of Rayleigh scattering is symmetric [2]. Fig. 4 shows the calculated scattering phase functions where the incident phonon is from the low-lying transverse acoustic branch (TA1) and the phonon states after scattering remain in the TA1 branch. Indeed, when the wavelength is even twelve times of the lattice constant $a$, the scattering phase functions for PbTe and GeTe are highly asymmetric, while it is symmetric for Si for the same wavelength. As the wavelength decreases, the scattering phase function becomes more asymmetric for all materials. However, for Si, the asymmetry exists only at very short wavelength ($\leq$ ~2a). This behavior is not limited to the TA1-to-TA1 scattering channel shown in Fig. 4 but also exists for all the other acoustic-to-acoustic scattering channels. The scattering phase functions for TA1-to-longitudinal acoustic branch (LA) and LA-to-LA are shown in Fig. S3 and S4 in the Supplementary Material. These observations indicate that the strain field is a much more important contributor to phonon scattering than a local empty site in PbTe and GeTe. The strain field induced by a point defect in these materials is spatially extended to cause the Mie scattering rather than the Rayleigh scattering down to 0.1 THz. This is similar to the Mie scattering of phonons by a finite-sized nanoparticle inclusion [53-56]. Rayleigh scattering therefore can only happen at lower frequencies or longer wavelengths, for example, < 0.1 THz for the TA branch of PbTe, as shown in Fig. S5 in the Supplementary Material.

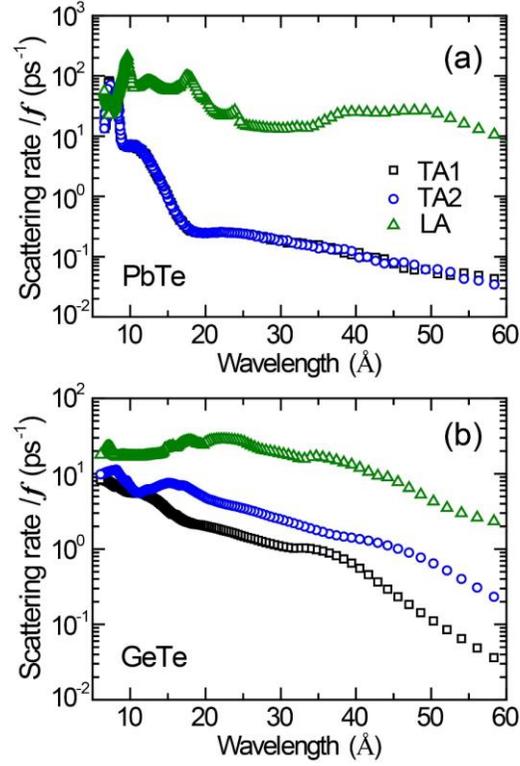

Fig. 3. Phonon-vacancy scattering rates normalized by the atomic fraction of vacancy as a function of phonon wavelength for (a) PbTe (Pb vacancy) and (b) GeTe (Ge vacancy) along the $\Gamma$-X direction. The scattering rates are calculated by the **T**-matrix approach. The TA1 refers to low-lying transverse acoustic branch and LA refers to longitudinal acoustic branch.

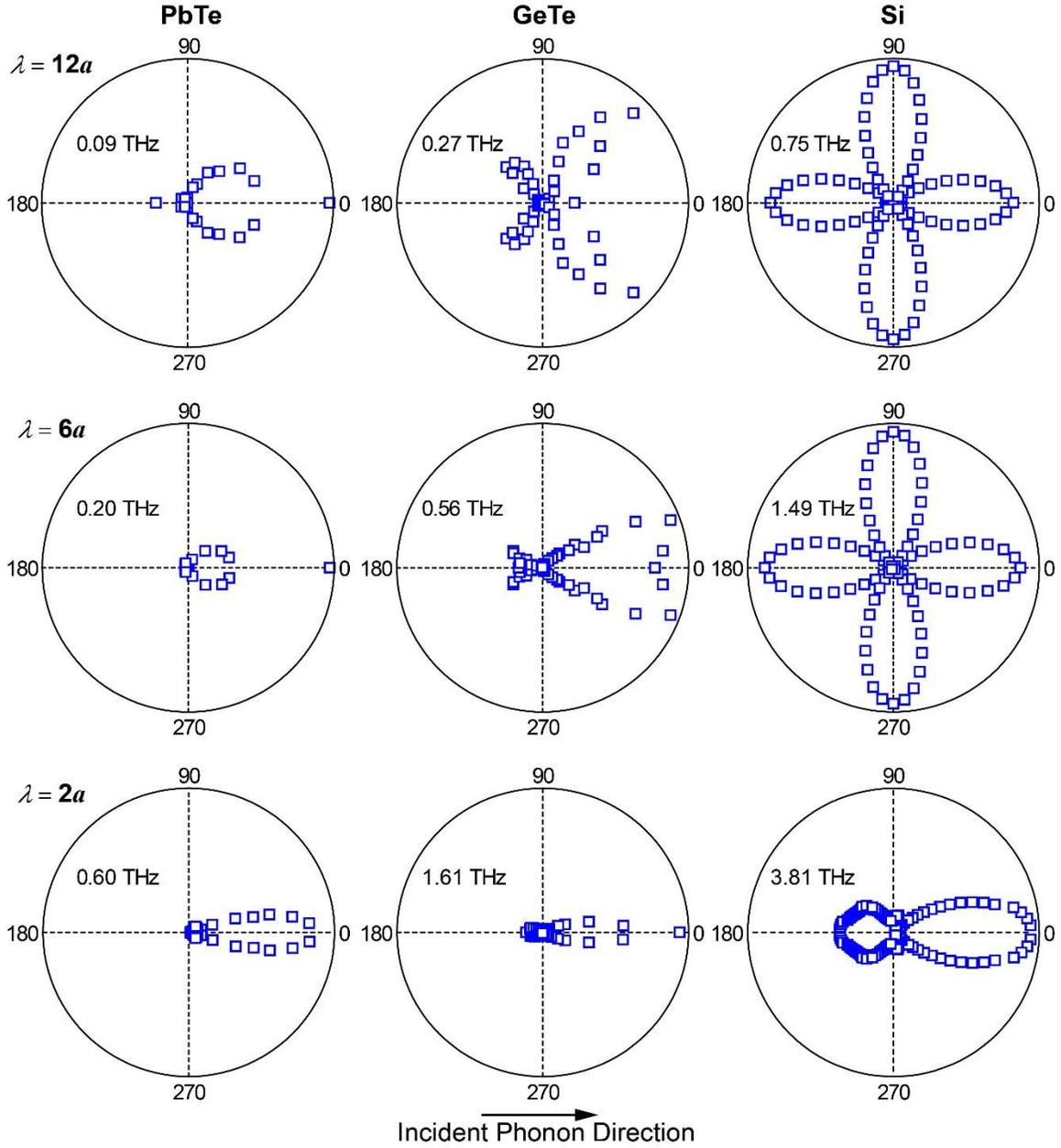

Fig. 4. Scattering phase functions by a vacancy in PbTe (Pb vacancy), GeTe (Ge vacancy), and Si, where phonon states before and after are in the TA1 branch. The incident phonon along the [001] direction has an incident angle of 0 degree and a wavelength of twelve, six, and two times of the lattice constant *a* for each material, respectively.

The asymmetric scattering phase function shows that phonon transport cannot be described with the simple kinetic theory of phonon gas that assumes each scattering process attempts to relax the phonon distribution into the symmetric equilibrium distribution as in the commonly used RTA. The RTA is exact when the scattering phase function is symmetric as shown in the Supplementary Material; all off-diagonal elements of a scattering matrix are canceled out and thus the relaxation of phonons at a certain mode can be described independently from other modes. However, when the scattering phase function is asymmetric, the off-diagonal elements of a scattering matrix become important and thus the relaxation of phonons at a certain mode is coupled to other modes. A similar behavior has been shown in the high thermal conductivity materials where normal phonon-phonon scattering is much stronger than umklapp phonon-phonon scattering; the strong normal scattering results in the asymmetric scattering phase function because of the phonon momentum conservation [5].

In Fig. 5, we calculate the lattice thermal conductivity values $\kappa_L$ of PbTe, GeTe, and Si with vacancy using RTA and the full scattering matrix. We choose a specific vacancy fraction that reduces $\kappa_L$ by ~50% at 300 K such that the phonon-vacancy scattering is comparable to the phonon-phonon scattering in terms of the strength for each material. The vacancy fraction for GeTe and PbTe is 1%, which is practically relevant as many previous studies report this vacancy level even for single crystals [25,34,35]. Fig. 5 shows that the RTA for the phonon-vacancy scattering substantially underestimates $\kappa_L$ of PbTe and GeTe but produces almost identical results as the full scattering matrix for Si. At 300 K, the RTA underestimates $\kappa_L$ by 15 to 20% for PbTe and GeTe when the vacancy fraction is 1%.

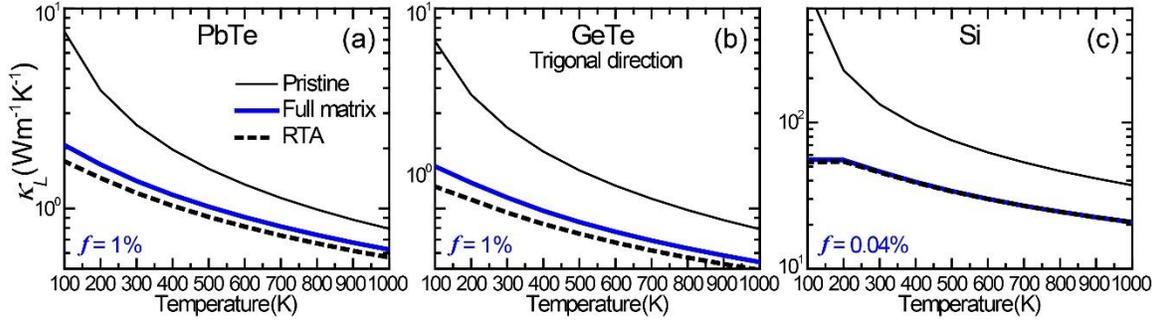

Fig. 5. Temperature dependent lattice thermal conductivity of (a) PbTe, (b) GeTe (trigonal direction), and (c) Si. The $f$ is atomic vacancy fraction. For the RTA case, only diagonal terms of phonon-vacancy scattering matrix are included while the full matrix case includes all elements of the scattering matrix. The phonon-phonon scattering for both cases is included with a full matrix. The thermal conductivity of GeTe along the binary direction is similar to the trigonal direction case and is shown in Fig. S6 in the Supplementary Material.

We further analyze the effects of Mie scattering on phonon spectral thermal conductivity. In Fig. 6, we show the spectral thermal conductivity contribution calculated by RTA and the full scattering matrix, providing a simple measure of asymmetricity of scattering phase function and the extent of Mie scattering in the phonon frequency space. From Fig. 6, the RTA and full scattering matrix solutions for PbTe and GeTe show a noticeable difference in a wide range of phonon spectrum, indicating that the Mie scattering is a better description than the Rayleigh scattering in those materials for the wide range of phonon spectrum. This is consistent with the observations in scattering phase functions. Fig. 4 shows that the scattering phase function of PbTe and GeTe is asymmetric at very low phonon frequencies for thermal transport (0.09 THz for PbTe and 0.27 THz for GeTe) and become even more asymmetric as the frequency increases. However, the marked difference between the RTA and full scattering matrix solutions for Si exists only around 5 THz and 12 THz which correspond to the near zone boundary states of transverse and longitudinal acoustic phonons, respectively, where Mie scattering occurs because of extremely short phonon wavelengths. Most phonon states contributing to thermal transport experience the Rayleigh scattering in Si.

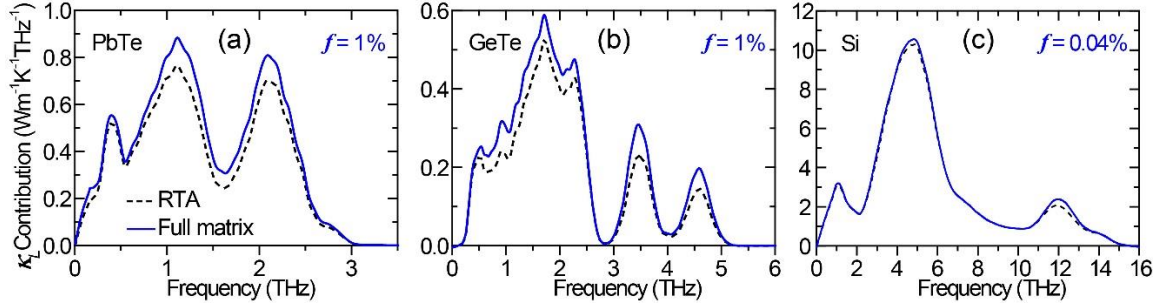

Fig. 6. Spectral lattice thermal conductivity of (a) PbTe, (b) GeTe (trigonal direction), and (c) Si, calculated by the full matrix and RTA at 300 K. The *f* is atomic vacancy fraction.

The Mie scattering behavior not only is limited to phonon-vacancy scattering, but also can be observed in phonon-dopant scattering. In Fig. 7, we present the scattering rates of phonons by two widely used dopants for PbTe, Na, and Bi substituting Pb [27-29]. The atomic fraction of dopants is usually high ranging from 1 to 2% for high thermoelectric figure-of-merits [27-29] and therefore its impact on the phonon scattering and lattice thermal conductivity can be significant. Fig. 7 shows that the scattering rate has much weaker frequency dependence than $\omega^4$, indicating Mie scattering by Na and Bi dopants similar to the Pb vacancy case we have discussed. In addition, the scattering phase function shown in Fig. S7 in the Supplementary Material is highly asymmetric for both dopants, confirming the simple kinetic theory can significantly overestimate the thermal resistance by dopants.

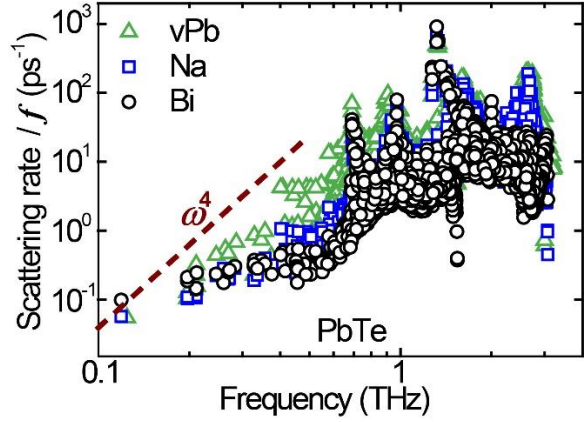

Fig. 7. Comparison of phonon scattering rates by Pb vacancy (vPb), Na and Bi substitution for Pb in PbTe from the **T**-matrix approach. The scattering rates are normalized by the atomic fraction of point defects.

## 4. Conclusion

In summary, we have applied an *ab initio* Green's function approach to investigate the scattering of phonons by point defects in PbTe, GeTe, compared with Si. In contrast to the often assumed Rayleigh scattering, the scattering of most acoustic phonons in PbTe and GeTe exhibits Mie scattering behaviors because of the long-range strain field induced by point defects. Specifically, the frequency dependence of scattering rates is weaker than the $\omega^4$ of Rayleigh scattering and the scattering phase function is highly asymmetric. The latter results in the failure of the relaxation time approximation, with an error of ~20% in predicting the lattice thermal conductivity at 300 K when the vacancy fraction is 1%. Our results highlight the importance of the long-range strain field induced by point defects for thermal transport. Similar behavior is expected in other materials with high electronic polarizability (such as IV–VI, $V_2$–$VI_3$, and elemental V materials) and common ferroelectric materials (such as perovskite oxides), which can lead to the long-range strain field by a point defect. The significant Mie scattering behavior shows the complexity of phonon scattering by point defects in IV-VI semiconductors, which cannot be described by the simple kinetic theory combined with the conventional Rayleigh scattering.

**Acknowledgements**

The authors thank Nebil A. Katcho and Lucas Lindsay for helpful discussions. The authors acknowledge support from National Science Foundation (Award No. 1705756 and 1709307). This work used the Extreme Science and Engineering Discovery Environment (XSEDE) Linux cluster at the Pittsburgh Supercomputing Center through Allocation No. TG-CTS180043 and the Linux cluster of the Center for Research Computing at the University of Pittsburgh.

**Supplementary Material for**

**Mie Scattering of Phonons by Point Defects in IV-VI Semiconductors PbTe and GeTe**

Ruiqiang Guo[1] and Sangyeop Lee[1,2,*]

[1]*Department of Mechanical Engineering and Materials Science, University of Pittsburgh, Pittsburgh, Pennsylvania 15261, USA*

[2]*Department of Physics and Astronomy, University of Pittsburgh, Pittsburgh, Pennsylvania 15261, USA*

*sylee@pitt.edu

# I. Supplementary Table

Table S1. Parameters for the calculations: i) cutoff energy of plane wave basis and the sampling of electronic states (**k**-mesh) in the first Brillouin zone for the DFT calculation; ii) the lattice constants from the density functional theory calculation; iii) the sampling of phonon states in the first Brillouin zone (**q**-mesh) for thermal conductivity calculation.

|      | Cutoff energy (eV) | **k**-mesh | Relaxed lattice constant (Å) | **q**-mesh |
|------|--------------------|------------|------------------------------|------------|
| PbTe | 500                | 15×15×15   | 6.57                         | 16×16×16   |
| GeTe | 500                | 15×15×15   | 6.18                         | 16×16×16   |
| Si   | 500                | 15×15×15   | 5.47                         | 24×24×24   |

## II. Supplementary Figures

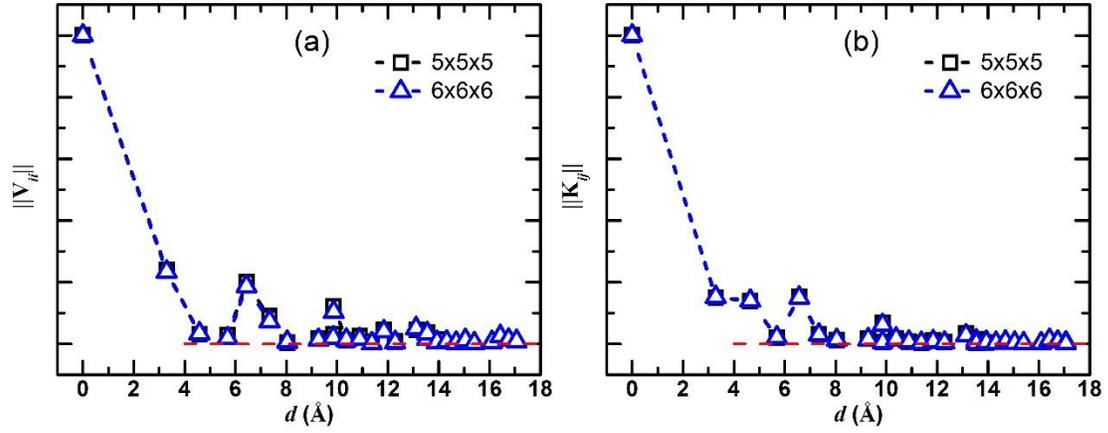

Fig. S1. Convergence test of IFCs in PbTe: (a) Frobenius norm of the change in self-interaction IFCs as a function of distance from the vacancy site. (b) Frobenius norm of IFCs in perfect crystals as a function of the interatomic distance. Note that all the values are normalized by the norm at the vacancy site.

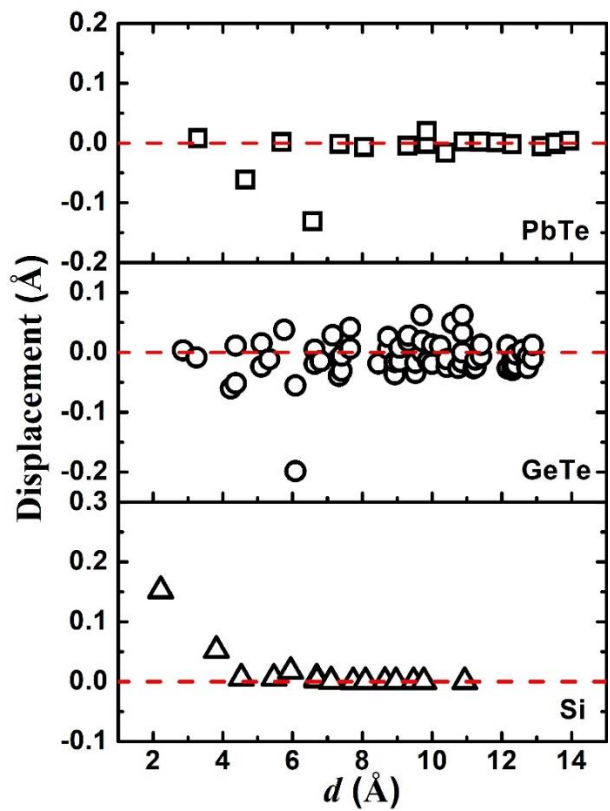

Fig. S2. Displacement of atoms as a function of the distance from the vacancy site. Data points above the dashed line indicate the increase of the distance from the vacancy site.

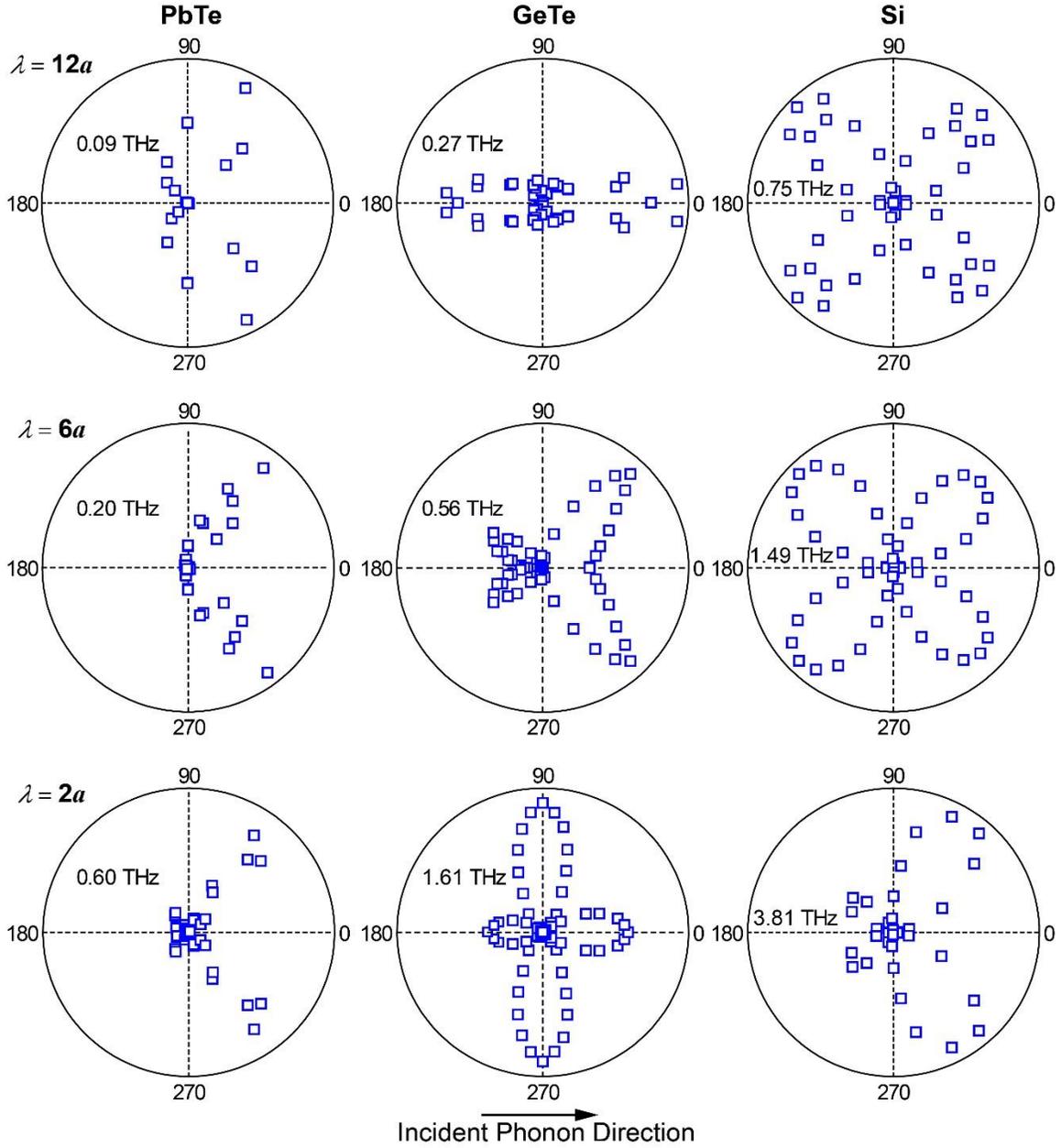

Fig. S3. Scattering phase functions by single vacancy in PbTe, GeTe and Si, where the incident phonon from the low-lying transverse acoustic (TA1) branch is scattered into the longitudinal acoustic (LA) branch. The incident phonon along the [001] direction has an incident angle of 0 degree and a wavelength of twelve, six and two times of the lattice constant $a$ for each material, respectively.

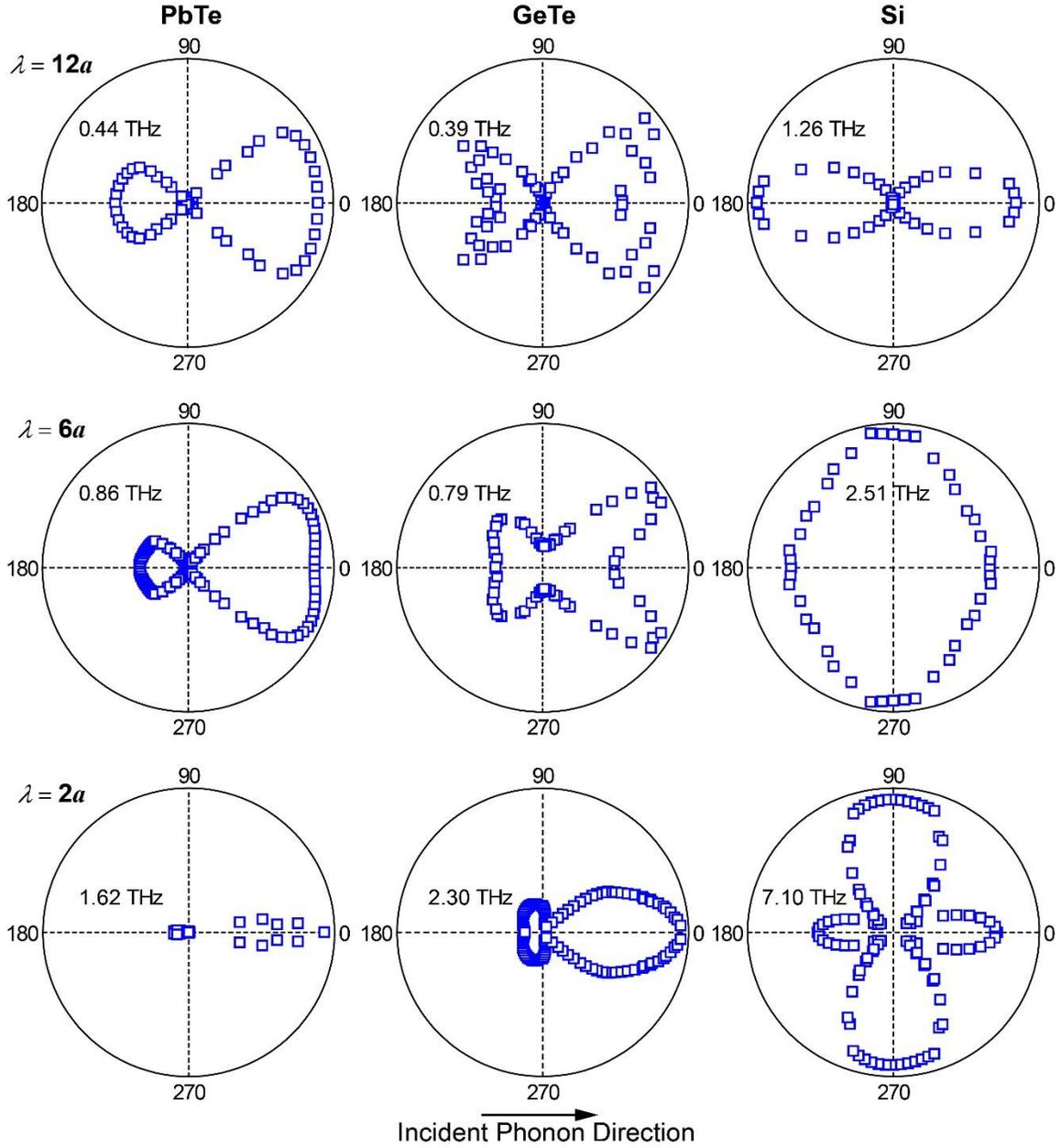

Fig S4. Scattering phase functions by a vacancy in PbTe, GeTe, and Si, where the phonon states before and after the scattering are from the LA branch. The incident phonon along the [001] direction has an incident angle of 0 degree and a wavelength of twelve, six and two times of the lattice constant $a$ for each material, respectively.

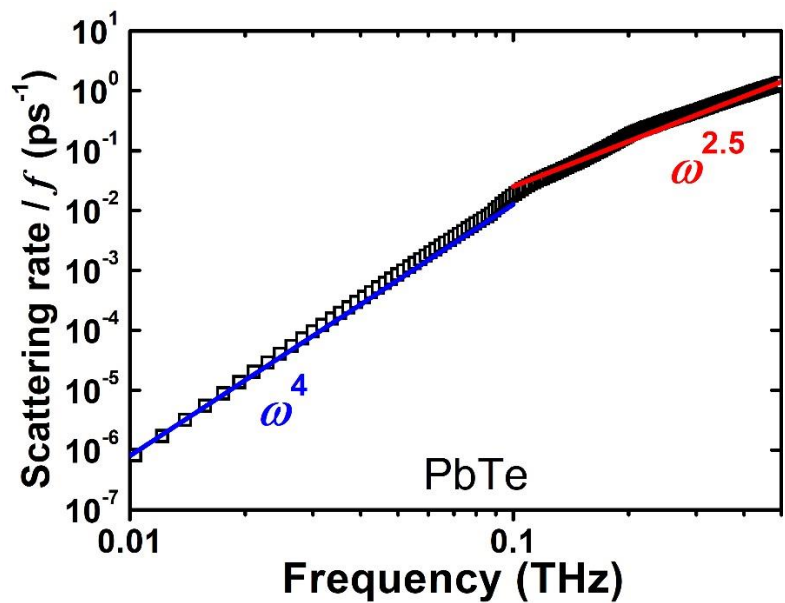

Fig. S5. Phonon-vacancy scattering rates normalized by the atomic fraction of vacancy of the TA branch along the Γ-X for PbTe (Pb vacancy) using **T**-matrix.

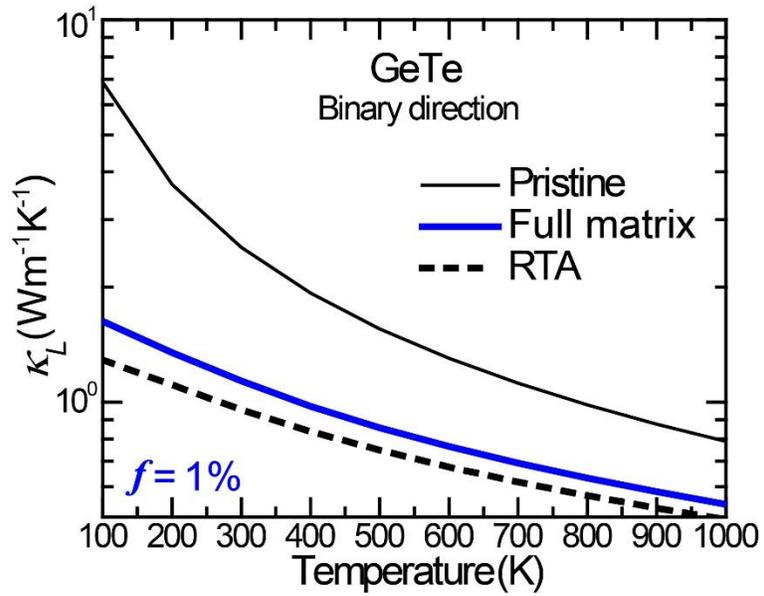

Fig. S6. Temperature dependent lattice thermal conductivity of perfect crystalline GeTe and GeTe with vacancy (the atomic fraction $f$ is 1%) along the binary direction. For the RTA case, only diagonal terms of phonon-vacancy scattering matrix are included while the full matrix case includes all elements of the scattering matrix. The phonon-phonon scattering for both cases is included with a full matrix.

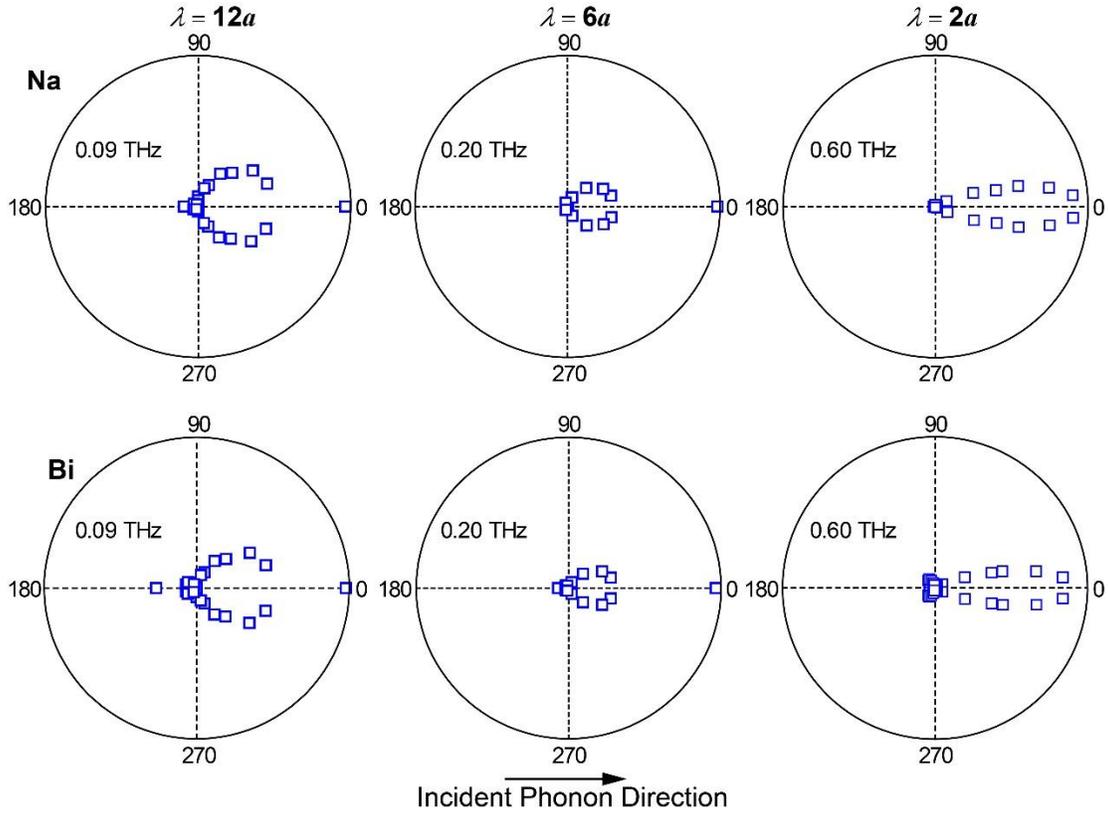

Fig. S7. Scattering phase functions by Na and Bi substitution of Pb for incident phonons along the [001] direction with various wavelengths in PbTe. Both incident and scattered phonons are from the TA1 branch. The incident phonon propagates with an angle of 0 degree.

# III. Supplementary Derivations

Here we show that the relaxation time approximation is exact only when the scattering phase function is symmetric. Starting with the Peierls-Boltzmann transport equation for an infinitely large sample with a spatially homogenous temperature gradient along *x*-direction:

$$-v_x(\mathbf{q}s)\frac{\partial f_0(\mathbf{q}s)}{\partial T}\frac{dT}{dx} = \sum_{\mathbf{q}'s'}[f(\mathbf{q}'s')S(\mathbf{q}'s',\mathbf{q}s) - f(\mathbf{q}s)S(\mathbf{q}s,\mathbf{q}'s')] \:, \quad (S1)$$

where $S(\mathbf{q}s,\mathbf{q}'s')$ is the transition rate from a phonon mode $\mathbf{q}s$ to another phonon mode $\mathbf{q}'s'$. The phonon distribution function $f$ can be decomposed into the equilibrium part $f_0$ and an asymmetric part $f_A$ displaced by the temperature gradient,

$$f(\mathbf{q}s) = f_0(\mathbf{q}s) + f_A(\mathbf{q}s) \:. \quad (S2)$$

With the detailed balance, the scattering term in Eq. (S1) can be written as

$$\sum_{\mathbf{q}'s'}[f_A(\mathbf{q}'s')S(\mathbf{q}'s',\mathbf{q}s) - f_A(\mathbf{q}s)S(\mathbf{q}s,\mathbf{q}'s')] \quad (S3)$$

Considering the relation $f_A(-\mathbf{q}s) = -f_A(\mathbf{q}s)$ in the homogenous temperature gradient, Eq. (S3) becomes

$$-f_A(\mathbf{q}s)\sum_{\mathbf{q}'s'}S(\mathbf{q}s,\mathbf{q}'s') + \sum_{\mathbf{q}'s'|q_x>0}f_A(\mathbf{q}'s')[S(\mathbf{q}'s',\mathbf{q}s) - S(-\mathbf{q}'s',\mathbf{q}s)] \quad (S4)$$

If the scattering phase function is symmetric, i.e., $S(\mathbf{q}'s',\mathbf{q}s) = S(-\mathbf{q}'s',\mathbf{q}s)$, the scattering integral can be reduced to a simple scattering term with a relaxation time:

$$\sum_{\mathbf{q}'s'}[f(\mathbf{q}'s')S(\mathbf{q}'s',\mathbf{q}s) - f(\mathbf{q}s)S(\mathbf{q}s,\mathbf{q}'s')] = -\frac{f_A(\mathbf{q}s)}{\tau} \:, \quad (S5)$$

where $\tau^{-1} = \sum_{\mathbf{q}'s'}S(\mathbf{q}s,\mathbf{q}'s')$.